\documentclass[letterpaper,12point,epsfig]{article}
\usepackage{epsfig}

\begin{document}

\title{Determining the Critical Temperature and Number of Frozen Layers in a Two-Dimensional Bed of Vibrating Hard Spheres Using a Global Equation of State}
\author{Alison E. Koser and Paul V. Quinn Sr. \\ Department of Physics, Kutztown University of Pennsylvania}
\maketitle

\begin{abstract}
Using a global equation of state, empirically derived by Luding, we accurately model the density profile of a two-dimensional hard sphere system with diameter $D$ and mass $m$ under gravity with a given temperature $T$ [Physica A, 271, 192 (1999)]. We then compare our solutions to MD simulated data. From the density profile, we can then solve for the critical temperature $T_c$, which we define as the temperature at which the system begins to condensate. Then, if $T$ is below $T_c$, there is some frozen portion of the system. We derive a formula for the number of frozen layers $\mu_f$, and compare our solution to the number of frozen layers in our simulated data.
\vspace{.4 cm}

\noindent PACS number(s) 64.70.Dv, 05.20.Dd, 51.10.+y

\end{abstract}

\section{Introduction} 

\hspace{.5 cm} Because of the large number of relatively small particles that compose a granular system, granular media tends to act like a collection of microscopic molecules.  However, individual grains in a granular system have a macroscopic mass.  Thus, particle collisions and gravity play a much larger role than in molecular systems, making it impossible to fully explain their mechanics in terms of kinetic theory alone.  Understanding hard sphere (HS) systems is crucial for analyzing important physical phenomena in our world, such as avalanches, earthquakes, and powder production. Newer models that correct for characteristics such as higher densities and dissipation in granular materials have succeeded in explaining simulated and experimental results that were previously left unsolved.

In a paper by Hong, [1], a method for modeling 
a granular system was developed by observing the effect of lowering the 
excitation level of a highly excited system of grains.  The excitation of grains is modeled as a temperature reservoir in contact with the system. When the kinetic temperature, $
T = \frac{1}{2}m \langle v^2 \rangle$, is high, particles in the system 
behave like a gas, where the mean free path is much larger than the particle size. The particles in this state are free to 
exchange positions with neighboring particles. At the other extreme, $T=0$, the particles remain motionless at the bottom of the 
container.  In this state, the particles behave like a 
solid, remaining locked in a cage, not free to exchange positions with any neighboring particles. As the temperature is lowered from the gas-like state, we note that a portion of the particles begins to move to the bottom, forming a solid regime, where a layer of particles is locked in a cage.  As the temperature gets closer to $T=0$, the number of particles condensing to the bottom increases, causing the solid regime, and hence, the number of condensed layers, to grow. This 
observation lead to the derivation of a critical temperature, $T_c$, at which the solid regime begins to form. Hong was able to determine this critical temperature as a function of the control parameters of 
the system. He also showed that there is a way to predict 
the solid regime as a function of $T$ once it has dropped below the critical temperature, $T_c$.  In another paper by Quinn et al [2], the theory proposed by Hong was tested and verified using an event driven(ED) molecular dynamics simulation code. The results of the ED simulations were compared with the proposed theory.  Hong derived results in both two and three dimensions, choosing a particular correlation function for each.  In two dimensions, Hong chose to use the correlation function proposed by Ree and Hoover [3], expressed as follows:
$$\chi(\phi)=\frac{1-\alpha_1\phi+\alpha_2\phi^2}{(1-\alpha\phi)^2},$$ 
with $\alpha = 0.489351\cdot(\pi/2)\approx 0.76867$, $\alpha_1=0.196703
\cdot(\pi/2)\approx 0.30898$, and $\alpha_2=0.006519\cdot(\pi^2/4)
\approx 0.0168084$.  Using this correlation function, Hong derived the following expression, 
$$-\beta(\zeta-\bar\mu) = \ln\phi + c_1\phi + c_2\ln(1-\alpha\phi)
+ \frac{c_3}{1-\alpha\phi} + \frac{c_4}{(1-\alpha\phi)^2}$$
with
$$ \beta\bar\mu = \ln\phi_0 + c_1\phi_0 + c_2\ln(1-\alpha\phi_0)
+\frac{c_3}{1-\alpha_o} + \frac{c_4}{(1-\alpha\phi_0)^2}$$
where 
$$c_1 = \frac{2\alpha_2}{\alpha^2}\frac{\pi}{2} \approx 0.0855,$$
$$c_2 = -\frac{\pi}{2}\frac{1}{\alpha^2}\left(\alpha_1 - \frac{2\alpha_2}{\alpha}\right)\approx -0.710,$$
$$c_3=-c_2 \approx 0.710,$$ 
$$c_4= \frac{\pi}{2}\frac{1}{\alpha}\left(1-\frac{\alpha_1}{\alpha} + 
\frac{\alpha_2}{\alpha^2}\right)\approx 1.278,$$
and $\phi_0$ is the density at $\zeta=0$.   
Furthermore, Hong was able to determine the following expressions for the critical temperature $T_c$,
$$T_c= \frac{mgD\mu\phi_0}{\mu_0}$$
and the number of frozen layers $\zeta_f$,
$$\zeta_f = \left(\mu-\frac{\mu_0}{\beta\phi_0}\right) = 
\mu \left(1-\frac{T}{T_c}\right).$$

The theory proposed by Hong broke the system of hard spheres into two regions, a solid region and a liquid region.  The density profile proposed by Hong was shown by Quinn to only be valid for the liquid region of the system.  This is because the correlation function proposed by Ree and Hoover, was not valid for a high density region of hard spheres.  Therefore it makes sense that the density function derived from this correlation function, would be inadequate for modeling the high density of the solid region.  In a paper by Luding et al, [4] a new correlation function was proposed for a two dimensional system.  This correlation function is valid for both high and low density regions of a system of hard spheres.  In this paper, we will derive a new density profile using the correlation function proposed by Luding, and compare it to the data produced by event driven molecular dynamics simulations.  We will also derive expressions for the critical temperature, $T_c$, and the number of frozen layers, $\zeta_f$.  We will also compare these expressions to the simulated data.

\section{Liquid-Solid Transition for Hard-Sphere Systems}
\hspace{5mm}  In his paper, Hong started with the Enskog Equation for hard spheres [5], 
$$\frac{\partial f}{\partial t} + {\bf v}\cdot {\bf \nabla f} - 
mg\frac{\partial f}{\partial v_z} = J_E,$$ 
where the Enskog collisional operator, $J_E$, is given by
$$J_E = D^2 \int d^3{\bf v_1}\int_+ d^2{\mathrm {e}}({\bf\hat{e}}\cdot{\bf g})
[f({\bf r},{\bf v'}) f({\bf r}+D{\bf \hat{e}},{\bf v_1'}) \chi
({\bf r}+D{\bf \hat{e}}/2)$$
$$- f({\bf r},{\bf v}) f({\bf r}-D{\bf \hat{e}},
{\bf v_1}) \chi({\bf r}-D{\bf \hat{e}}/2)].$$
From this equation, he derived the density profile used to describe the liquid region of the hard sphere system.  The same expression can be derived starting with the pressure difference equation, taught in most introductory physics classes.  The pressure difference equation that describes the change of pressure \textit{P} over height \textit{z}, in terms of gravity \textit{g} and the density $\rho$, 
$$\frac{dP}{dz}=-\rho g \eqno(1) $$
As a simple example, we can show how this simple equation can be used to derive the Maxwell-Boltzmann Distribution.  Starting with the Ideal Gas Law,
$$PV = nkT,$$
where \textit{k} is Boltzmann's constant and \textit{T} is the temperature of the system, one can get the following Ideal Gas Pressure in terms of the density, $\rho$:
$$P = \rho k T$$
Substituting this expression into Eq.(1) and solving for $\rho$, we obtain the following density profile,
$$\rho=\rho_o exp\Big(\frac{-gz}{kT}\Big).$$
This is the Maxwell-Boltzmann Distribution, used to model a gaseous phase in statistical mechanics.  Since the function only applies for the gaseous phase, it does not accurately fit the liquid-solid transition in our system of hard spheres.

From the kinetic theory, we can relate pressure and density, where \textit{m} equals the mass of a single particle,
$$P=\frac{T}{m}[\rho +2\rho \chi (\rho)].$$
$\chi (\rho)$ is a correlation function that takes into account the probability of a collision occurring among particles.

In 2001, Stephan Luding [4] proposed the following Global Pressure Equation to combine the low-density pressure and high-density expression:
$$P=\rho+\rho P_4+\rho m(\rho)[P_{dense}-P_4] \eqno(2)$$
where
$$P_4 = =\frac{T}{m}[\rho +2\rho \chi_4 (\rho)],$$
 
$$\chi_4(\rho)=\frac{1-\frac{7\rho}{16}}{(1-\rho)^2}-\frac{\rho^{\frac{3}{16}}}{8(1-\rho)^4},\eqno(3)$$
and
$$P_{dense}=\frac{C_o}{\rho_{max}-\rho}[1+C_1(\rho_{max}-\rho)+C_3(\rho_{max}-\rho)^3] -1$$

$\chi_4$ is a correlation function, similar to the form proposed by Ree and Hoover. The first term in Eq.(3) is a simpler correlation function,
introduced by Henderson [6,7].  The correlation function is determined via a virial expansion around low densities.  The value of
$P_4$, taken at contact, accounts for the excluded volume
effect and the increase of the collision rate with density.  
$P_{dense}$ is Luding's proposed corrected high-density pressure with the numerically fitted constants $C_o\approx 1.8137$, $C_1=-.04$, and $C_3=3.25$.

Included in the Global Pressure Equation, is the following empirical merging function,
$$m(\rho)=\frac{1}{1+exp[\frac{\rho_c-\rho}{m_o}]}$$
with the numerical fitting constant $m_o=0.0111$. Unfortunately, this function inhibits our ability to analytically solve for the density function $\rho$.  We can obtain an integral expression for the density profile by substituting Eq.(2) into the pressure difference equation, Eq.(1), and solving for the position $z$.  We obtain:

$$z =\frac{-T}{mg} \int_{\rho_{o}}^{\rho} [ \frac{1}{\rho}+2\rho+ \frac{C_o m(\rho)}{(\rho_{max}-\rho)^2}+\frac{C_o}{(\rho_{max}-\rho)} \frac{dm(\rho)}{dz} +\frac{C_o m(\rho)}{\rho(\rho_{max}-\rho)}+$$

$$2C_oC_3(\rho_{max}-\rho)m(\rho) +\frac{C_oC_3}{(\rho_{max}-\rho)} \frac{dm(\rho)}{dz}+\frac{C_oC_3m(\rho)}{\rho(\rho_{max}-\rho)^2}+$$

$$\frac{(C_oC_1-1)m(\rho)}{\rho}-2\rho m(\rho)\frac{dP_4}{dz}-4m(\rho)\frac{dP_4}{dz}-4\rho P_4\frac{dm(\rho)}{dz} d\rho]+z_o \eqno(5)$$

Unfortunately, there is no analytical solution to Eq.(5), but using numerical methods, one can obtain the solution, and hence, a single function to completely describe the density profile of a two dimensional system of hard spheres under gravity.

\section{The Critical Temperature and Number of Frozen Layers}

One can define the critical temperature as the point at which the density at the bottom layer, $\rho_o$, becomes the closed packed density $\rho_c$ such that
$$\rho_o (T_c)=\rho_c.$$
For $T\leq T_c$, a portion of the particles near the bottom settle into their minimum energy positions.  The particles in this region are essentially locked in a small cage, where they are free to slightly wiggle, but are never able to exchange positions with other particles in the system.  Hence, they form a crystal-like structure.  

The maximum density $\rho_o$ depends on the underlying crystalline structure. The density of square lattice packing is $\rho=\frac{\pi}{4}$, and the density of a hexagonal lattice is $\rho=\frac{\pi \sqrt{3}}{6}$. We say that particles become locked in position when the density reaches the square-packed density $\rho_c=\frac{\pi}{4}$. However, we will see that the lattice of the solid regime can fluctuate between square and hexagonal packing.

We can use Eq.(5) and particle conservation to derive an expression for the temperature, $T_c$ at which the solid region begins to form at the bottom.  We begin with the conservation of particles in a column which can be expressed as follows:

$$\int_{\rho_o}^{0} z(\rho)d\rho = h, \eqno 6 $$

\noindent where $h$ is the height of the column of balls at a particular temperature, $T$.  The height $h$ can be rewritten as the number of layers simply by factoring out the diameter of a particle, $D$, such that

$$\mu = \frac{h}{D}.$$

This allows us to rewrite Eq.(6) as 

$$\mu = \frac{1}{D} \int_{\rho_o}^{0} z(\rho)d\rho.  \eqno(7)$$  

The variable $\beta$ can be defined as 

$$\beta = \frac{mg}{T}$$

\noindent We can now define the constant

$$\beta_c = \frac{mg}{T_c},$$

\noindent where $T_c$ is the critical temperature and the constant $h_o$ such that

$$h = D \mu=\frac{h_o}{T_c}.$$

Now Eq.(5) can be rewritten as

$$ -\beta_c z = I(\rho)-I(\rho_o), \eqno(8)$$

\noindent where the function $I(\rho)$ comes from the upper limit of 
the integrals in Eq.(5), and $I(\rho_o)$ comes from the lower limit of the integrals in Eq.(5).  Then we can rewrite Eq.(8) as 

$$-\beta_c z =I(\rho)-\beta \bar{h}, \eqno(9)$$

\noindent where $\beta\bar{h}=I(\rho_o)$.

\noindent Solving Eq.(9) for $z$ yields

$$z = -\frac{1}{\beta_c}[I(\rho) - \beta_c \bar{h}]. \eqno(10)$$

We can now substitute Eq.(10) into Eq.(7) to obtain 

$$\mu = \frac{1}{\beta_c D} \int_{\rho_o}^{0} z(\rho)d\rho$$ 
$$= \frac{-1}{\beta_c D}\int_{\rho_o}^{0}[I(\rho)-\beta_c \bar{h}]d\rho$$
$$=\frac{-1}{\beta_c D}[\int_{\rho_o}^{0}[I(\rho)d\rho - \beta_c \bar{h} \rho]$$
$$=\frac{1}{\beta_c D}\int_{0}^{\rho_o}[I(\rho)d\rho - \beta_c \bar{h} \rho]. \eqno(11)$$

\noindent However, by our previous definitions, we can state that

$$\mu =\frac{h_o}{D \beta_c} = \frac{T_c h_o}{D m g}. \eqno (12) $$

\noindent Using Eq.(12), we obtain an expression for the critical temperature $T_c$,

$$T_C = \frac{mgD\mu}{h_o}, \eqno(13)$$

\noindent where $h_o$ is obtained using the following expression:

$$h_o = \int_{0}^{\rho_o} [I(\rho)-I(\rho_o)]d\rho.$$

Using the Global Equation of State and Eq.(5), we calculate the value of $h_o$ to be 26.8097.  Thus,

$$T_c = \frac{mgD\mu}{26.8097}. \eqno(14)$$

We can now derive an expression for the number of frozen layers, using the critical temperature from Eq.(14).  The number of frozen layers, 
$\zeta_f$ will simply be the total number of layers, $\mu$, less the 
number of fluidized layers on the top, $h_o/D \beta$.  This gives us the following expression for the number of solid layers:

$$\zeta_f = \mu - \frac{h_o}{D \beta} = \mu - \frac{\mu D \beta_c}{D \beta}. \eqno (15)$$

Finally we obtain the following expression for the number of layers by simplifying Eq.(15):

$$\zeta_F = \mu(1 - \frac{T}{T_c}). \eqno(16)$$

The expression we have derived is exactly the same as that derived by Hong in [1], with a modified value of $T_c$, due to the Global Equation of State.

\section{Molecular Dynamics Simulations}

We have used an event driven(ED) molecular dynamics code to test the 
condensation of hard spheres due to gravity. We refer the readers to 
references [8,9,10] for details of the algorithm regarding the collisional dynamics. This particular code takes into account the rotation of the hard spheres as well as a way to handle the inelastic collapse. The difference between ED and soft-sphere molecular dynamics codes lies in the way time is advanced during the simulations. Instead of using finite time increments to advance the system, the ED algorithm finds and advances the system to the next possible event, usually a collision between particles or between a particle and a wall. This process advances the system in time, but by different time steps for each event that occurs. The thermal reservoir of our system was modeled using a white noise driving, first introduced by Williams 
and Mackintosh [11]. In this model, a random velocity kick, $\Delta 
\vec{v}$, is added to each particle's velocity every set interval of time, $\Delta t$. These random velocity kicks are not correlated with each other in any way. The program allows us to set the temperature of the system by setting an input parameter controlling the width of the Gaussian from which the random kicks are drawn. In effect, the temperature parameter controls the temperature of the reservoir and hence, the kinetic temperature of the system. Note that we are {\it not} driving the system with a bottom wall connected to a temperature reservoir, which is used fairly often as a model for the vibrating bed.

Fig.(1) displays data from a typical simulation of 1000 particles, with initial layers , $\mu=40$ layers, radii $r=.0001$ m, and mass $m=8.378 \times 10^{-9}$ kg. This data is fit with the three density functions derived previously, the Mazwell-Boltzmann Distribtuion, the Enskog approximation using $P_4$, and Eq.(5) derived from the Global Equation of State. It is clear that the Maxwell-Boltzmann Distribution only works for the very low density region, or the gaseous region as expected, and the Enskog approximation breaks down once the particles reach their maximum denisty, or solid-like region.  Eq.(5) fits the data for all densities, making the Global Equation of State the most accurate for representing our granular system.  

We now present typical configurations as a function of temperature using Eq.(5). In Figs.(2), (3) and (4) there are three representative samples at different temperatures for 1000 particles with initial layer numbers of $\mu = 40$, of mass $m=8.378 \times 10^{-9}$ kg and diameter $D=0.0002$ m. The initial layer thickness is $\mu=40$, and the gravitational constant $g=9.81$ m/s$^2$.  Seen in Fig.(2), at a high 
temperature ($T_1 > T_c$ and $\rho_o<\rho_c$), all the particles are fluidized and dynamically active. The density profile is fit using the density derived from the Global Pressure Equation, Eq.(5). Since the condensation picture is based on elastic hard spheres, we used both a completely elastic system, where one burst of energy is used to start the system, as well as coefficients of restitution that were close to unity in these simulations. We note, however, that it was shown in [4] that the 
presence of dissipation {\it does not} change the condensation picture. The critical temperature is just shifted linearly on the order of the coefficient of restitution. In our simulations, the coefficient of restitution for all particle collisions in our inelastic systems, whether with the walls or other hard spheres, gives an elasticity of $\epsilon=0.99$.

At a lower temperature ($T_2 < T_c$), particles begin to condense and form a solid near the bottom, as shown in Fig.(3). Note that the 
particles first form a loose square-packed lattice, as in Fig.(3), and then progressively evolve into a more compact hexagonal lattice structure, as shown in Fig.(4). Once the compact hexagonal solid is formed, the density at the bottom remains more or less constant at $\rho_o=\frac{\pi\sqrt{3}}{6}\approx .906$. This hexagonal structure is then, for the most part, permanently retained. For Fig. (3) and (4), the oscillations in the solid regime are real, but they are simply the finite size effect, i.e, the hexagonal packing in a finite lattice has two more particles in alternative layers. This oscillation would 
disappear in the thermodynamic limit.

The critical temperature $T_c$ is determined by the temperature at which a square-packed solid is formed at the bottom layer. Beyond this temperature, the solid regime steadily approaches the close-packed hexagonal structure and once attained is fixed at $\rho_o\approx .906$. We point out that between $T_c$, where there is a square-packed lattice, and lower temperatures, where there is a hexagonally packed lattice, particles squeeze themselves together, expelling holes and progressively forming a compact hexagonal solid. Recall that the critical temperature, Eq.(14), is a function of $\rho_0$ and $\mu_0$. Eq.(14) shows that $\mu_0$ itself is also a function of $\rho_0$. Therefore, as the system changes from square-packed to compact hexagonal packing, so do the values of $\rho_0$, $\mu_0$, 
and consequently $T_c$. This is because as $\rho_0$ increases as the solid gets more compact, causing $\mu_0$ to increase and $T_c$ to decrease. 

\section{Global Fits to Simulated Data}
\hspace{5mm} Using an Event-Driven Molecular Dynamic simulation written by Stephan Luding, we obtained data for height versus density for a two-dimensional vibrating bed of hard-discs. Table 1 shows what control parameters we used and how we varied them. For each parameter set, we ran several different simulations at different temperatures.

\begin{center}

Table 1: Simulation Parameters
\vspace{.4 cm}
  \begin{tabular}{|c|c| c | c  | c |}
    \hline
Parameter Set & Mass ($kg$) & Gravity $(\frac{m}{s^2})$& Diameter ($m$) & $\mu$ \\ \hline
    1 &  $8.378 \times 10^{-9}$ & 9.81 & .0002 & 20 \\ \hline
    2 &  $1.676 \times 10^{-9}$ & 9.81 & .0002 & 20 \\ \hline
    3 &  $8.378 \times 10^{-9}$ & 4.91 & .0002 & 20 \\ \hline
    4 &  $8.378 \times 10^{-9}$ & 9.81 & .0004 & 20 \\ \hline
    5 &  $8.378 \times 10^{-9}$ & 9.81 & .0002 & 40 \\ \hline
  \end{tabular}
\end{center}

Using the method of Reimann Sums in a computer program, we numerically calculated the global equation solution from Eq.(5). Then we fit the numerical solution for each density profile to obtain the temperature of each system. From Fig.(2), (3), and (4), it is clear that the numerical solution to Eq.(5) fits the data extremely well. We note that there are slight density oscillations in the solid regime for the simulated data due to the presence of the bottom wall.  

Fig.(2), (3) and (4) are all examples of completely elastic systems, with no energy dissipation.  Fig.(5) and (6) show comparisons between the elastic and inelastic trials. For the inelastic case, Fig. (6), it is clear that the Global equation still fits the density profile nicely in the liquid-solid regime. However, the density profile tends to underestimate the height at lower densities. This is because the particles in the fluid regime are active in the sense that they are free to move around, undergoing collisions, while those in the solid regime are confined in a cage, fluctuating around fixed crystalline positions, and do not undergo collisions. This effect is discussed further in [4], where the Global Equation was initially developed.

To calculate the number of frozen layers for each simulation, we use Eq. (16), where $T$ is calculated numerically in the simulation and $T_c$ is obtained using Eq.(14). We find, as expected, that the $\mu_f$ decreases with increasing temperature.  This can be seen in Fig.(7), which is data for the case of initial layer number $\mu = 40$, radii $r=0.0001 m$ and mass $m=8.378 \times 10^{-9}$ kg.  One can see from the linear fit that the data matches the predictions of Eq.(16).  The theoretical slope calculated from Eq.(16) is $-1.631 \times 10^{11}$ while the experimental slope obtained from the simulated data shown in Fig.(7) is $-1.668 \times 10^{11}$.  This is a percent difference of only $2.45\%$.  The other systems produced similar results, all within $5\%$ or less of the predicted values. 

Fig. (8) provides a snap-shot, where $T<T_c$. In this system, we would expect to find 30.95 frozen layers. Indeed, we can see from the picture that the first 31 layers appear frozen, whereas particles above this layer can move freely. These results illustrate validity of the predictions obtained by using the Global Equation of State.

\section{Conclusion}
\hspace{.5cm} There are points to consider due to the results presented in this paper.
First, we have demonstrated in this paper, that the Global Equation of State, first presented by Luding et al [4], does indeed account for the liquid-solid transition which exists in a system of hard spheres.  The profile derived from the equation of state is useful for determining the kinetic temperature of the system, from the configurational statistics.  The breakdown of particle conservation is no longer an issue when looking at the density profile, because the Global Equation of State, includes those particles frozen at the bottom.  However, we can still use the concept of a critical temperature to derive the number of frozen layers at the bottem, and compare it to what is observed in our simulated data, due to the fact that the density becomes constant as the solid regime becomes hexoganally-close packed.  The region that is close packed is considered the solid regime and can be used to determine the number of frozen layers in the system  
and in turn, the thickness of boundary layers.  Since only a fraction
of grains are mobilized under shear [12,13], and avalanches and
many interesting dynamics occur in these thin boundary layers [14,15],
such a determination should be of technological importance.

Second, the existence of a gravity induced liquid-solid transition of
hard spheres must have some interesting consequences to higher order
kinetic theory, in particular with regard to the dynamic
behaviors.  Unlike particles in the liquid regime,
those particles in the solid regime are largely confined in cages and fluctuate around fixed positions.  Their motions resemble
lattice vibrations rather than binary collisions, and it may be a little peculiar, albeit not unphysical, to attempt to
describe the lattice vibrations using kinetic theory. If so, 
such a description must include much more than binary collisions. However, the Global Equation of State gives a complete picture of a system in transition, accounting for both the liquid region with binary 
collisions as well as the solid region with the lattice regions.  The connection between the two regions, is an empirical one derived by Luding et al [4].  This gives us some insight into the kintetic details of the transition from liquid to solid regimes, and could be studied further.  

As discussed in the beginning and
demonstrated in this paper, this gravity induced liquid-solid 
transition is not a peculiar phenomenon
associated with the Global Equation of State, but rather an intrinsic transition inherent in a system where an excluded volume interaction is dominant.  The formation of a solid at the bottom is the appearance of a massive occupied low energy state due to the Pauli exclusion principle.  Therefore, the bridge between the liquid and solid regimes, as presented by the Global Equation of State and the upward spread of the solid regime {\it should} persist because the Pauli exclusion principle is in action in real space, even if one may use different approximations [16-25]  or may try a different form for the pressure, such as the form suggested by Percus-Yevick [26], and/or in higher order truncation. It only disappears in the limit when the
close volume packing density, $\rho$ becomes one, which is possible only in the case of an ideal Appolonian packing [27].  One should also further expand the ideas used to derive the Global Equation of State from two to three dimensions, to see if a similar equation can be derived and tested.  

Finally, we have shown that the presence of dissipation does not alter
the condensation picture at all [28,29], {\it as long as} the velocity distribution remains Gaussian.  Previous experiments [30] have demonstrated the non-Gaussian nature of the velocity distribution, but if the dissipation is small, which is the case for the inelastic simulations carried out in this work, the deviation from
Gaussian should be small.  We also point out that
for hard sphere systems without gravity, there exists no
typical energy scale, and thus any transition must be entropy driven,
i.e., there exists no critical temperature, and the
phase transition occurs at a critical volume fraction [31].  However, for the system considered in this paper, there exists a typical energy associated with the potential energy due to gravity, and thus this transition is not entropy driven, but energy driven, and therefore a critical temperature $T_c$ must exist.

\vskip 1.0 true cm
\noindent {\bf Acknowledgment}
We wish to thank Stefan Luding for providing us with his ED code and helpful discussions over the course of this work.

\section{References}

\noindent [1] D. C. Hong, {\it Physica A} {\bf 271}, 192 (1999).
\vspace{.4cm}

\noindent [2] Paul V. Quinn and D. C. Hong, {\it Physical Review E} {\bf 62} 8295(2000).
\vspace{.4cm}

\noindent [3] F. R. Ree and W. G. Hoover, {\it J. Chem. Phys. 40}, 939 (1964).
\vspace{.4cm}

\noindent [4] S. Luding, {\it Phys. Rev. E}, {\bf 63} 042201(2001).
\vspace{.4cm}

\noindent [5] D. Enskog and K. Sven, {\it Vetenskapsaked Handl.} {\bf 63}, 4 (1922).
\vspace{.4cm}

\noindent [6] D. Henderson, {\it Mol. Phys.} {\bf 30}, 971(1975).
\vspace{.4cm}

\noindent [7] D. Henderson, {\it Mol. Phys.} {\bf 34}, 301(1977).
\vspace{.4cm}

\noindent [8] B. D. Lubachevsky, {\it J. Comp. Phys.} {\bf 94}, 255(1991).
\vspace{.4cm}

\noindent [9] S. Luding, {\it Phys. Rev. E.} {\bf 52}, 4442 (1995).
\vspace{.4cm}

\noindent [10] S. Luding and S. McNamara, Granular Matter,1(3), 113 (1998).
\vspace{.4cm}

\noindent [11] D. M. Williams and F.C. MacKintosh, {\it Phys. Rev. E} {\bf 57}, R9 (1996).
\vspace{.4cm}

\noindent [12] D.M. Haynes, D. Inman, {\it J. Fluid. Mech.} {\bf 150}, 357 (1985).
\vspace{.4cm}

\noindent [13] L.E. Silbert et al, APS March Meeting Bulletin, P25.007 (2000).
\vspace{.4cm}

\noindent [14]  H. Jaeger, S.R. Nagel, and R. P. Behringer, {\it Rev. Mod. Phys.} {\bf 68}, 1259 (1996).
\vspace{.4cm}

\noindent [15] 'Granular Gases', Edited by S. Luding, Poschel,
H. Herrmann, Springer-Verlag (2000).
\vspace{.4cm}

\noindent [16] G. Rascon, L. Mederos, and G. Navascues, {\it Phys. Rev. Lett.}{\bf 77}, 2249 (1996). 
\vspace{.4cm}

\noindent [17] T. P. C. van Noije and M. H. Ernest, {\it Granular Matter}{\bf 1}, 57 (1998).  
\vspace{.4cm}

\noindent [18] B. Doliwa and A. Heuer, {\it Phys. Rev. Lett.}
{\bf 80}, 4915 (1988). 
\vspace{.4cm}

\noindent [19] S. Torquato, {\it Phys. Rev. E} {\bf 51}, 3170 (1995).
\vspace{.4cm}

\noindent [20] J. J. Brey, J. W. Dufty, C.S. Kim, and A. Santos, {\it Phys. Rev. E} {\bf 58}, 4638 (1998).
\vspace{.4cm}

\noindent [21] A. Santos, S. B. Yuste, and M.L.D. Haro, {\it Mol. Phys.} {\bf 96}, 1 (1999).  
\vspace{.4cm}

\noindent [22] P. Richard, L. Oger, J.-P. Troadec, and 
A. Gervois, {\it Phys. Rev. E} {\bf 60}, 4551 (1999). 
\vspace{.4cm}

\noindent [23] P. Sunthar and V. Kumaran, {\it Phys. Rev. E} {\bf 60},
1951 (1999).  
\vspace{.4cm}

\noindent [24] V. Kumaran, {\it Phys. Rev. E} {\bf 59}, 4188 (1999).
\vspace{.4cm}

\noindent [25] S. Luding, in {\it Granular Gases,} edited by
T. Poschel and S. Luding (Springer Verlag, Berlin, 2000).
\vspace{.4cm}

\noindent [26] J.K. Percus and G.J. Yevick, {\it Phys. Rev.} {\bf 110}, 1 (1958).
\vspace{.4cm}

\noindent [27] For Appolonian packing, see B. Mandelbrot, ``The Fractal
Geometry of nature,''(W. H. Freeman and Company, New York, 1982).
\vspace{.4cm}

\noindent [28] J.S. Olafsen and J.S. Urbach,
{\it Phys. Rev. E} {\bf 60}, R2468 (1999).
\vspace{.4cm}

\noindent [29] W. Losert, D. Cooper, D. Kudrolli, and J.P. Gollub, {\it Chaos}, {\bf 9}, 682 (1999).
\vspace{.4cm}

\noindent [30] D. C. Hong, Fermi Statistics and Condensation,
in 'Granular Gases,' edited by S. Luding and T. Poschel
Springer-Verlag (2000).
\vspace{.4cm}

\noindent [31] D. Frankel, {\it Physica A} {\bf 263}, 26 (1999).

\newpage
\Large
\noindent \bf Figure Captions
\normalsize
\normalfont

\vskip .4 true cm 

\noindent Figure 1:  This is a system of 1000 elastic two-dimensional spherical particles with initial layer number $\mu = 40$, radii $r=0.0001$ m and mass $m=8.378 \times 10^{-9}$ kg. This displays fits to the data using the Maxwell Boltzmann Distribution, the Enskog Equation using $\chi_4$, and Eq.(5), the density profile derived from the Global Equation of State.  

\vskip .2 true cm
\noindent Figure 2:  This is a system of 1000 elastic two-dimensional spherical particles with initial layer number $\mu = 40$, radii $r=0.0001 $m and mass $m=8.378 \times 10^{-9}$ kg. The temperature of this system is greater than $T_c$, hence all particles are fluidized. The red points represents the data points, while the blue is the density profile derived from the Global Equation of State.

\vskip .2 true cm

\noindent Figure 3:  This is a system of 1000 elastic two-dimensional spherical particles with initial layer number $\mu = 40$, radii $r=0.0001 $m and mass $m=8.378 \times 10^{-9}$ kg. The temperature of this system is slightly below $T_c$. The first few layers at the bottom are beginning to transition to a solid, while the majority of the particles are still fluidized. The red points represents the data points, while the blue is the density profile derived from the Global Equation of State.

\vskip .2 true cm

\noindent Figure 4:  This is a system of 1000 elastic two-dimensional spherical particles with initial layer number $\mu = 40$, radii $r=0.0001 $m and mass $m=8.378 \times 10^{-9}$ kg. The temperature of this system is much less than $T_c$. The large number of the bottom layers in the solid-like regime have become hexagonally packed, while a smaller number of particles remain fluidized. The red points represents the data points, while the blue is the density profile derived from the Global Equation of State.

\vskip .2 true cm

\noindent Figure 5:  This is a system of 1000 elastic two-dimensional spherical particles with initial layer number $\mu = 40$, radii $r=0.0001 $m and mass $m=8.378 \times 10^{-9}$ kg. The global fit for the density profile fits the elastic system of particles throughout the entire range of densities. The red points represents the data points, while the blue is the density profile derived from the Global Equation of State.

\vskip .2 true cm

\noindent Figure 6:  This is a system of 1000 inelastic two-dimensional spherical particles with initial layer number $\mu = 40$, radii $r=0.0001 $m and mass $m=8.378 \times 10^{-9}$ kg. The global fit for the inelastic density profile works well for high density but deviates at lower densities. The red points represents the data points, while the blue is the density profile derived from the Global Equation of State.

\vskip .2 true cm

\noindent Figure 7:  This is a graph of layer number versus temperature for a system of 1000 elastic two-dimensional spherical particles with initial layer number $\mu = 40$, radii $r=0.0001 $m and mass $m=8.378E-9 $kg.  As the temperature decreases, more particles become locked in the solid regime. One can see that the relationship is linear, as is predicted by Eq.(16).  

\vskip .2 true cm

\noindent Figure 8:  This is a snapshot from a system of 1000 elastic two-dimensional spherical particles with initial layer number $\mu = 40$, radii $r=0.0001 $m and mass $m=8.378 \times 10^{-9}$ kg. The temperature of this system is much less than $T_c$. This simulation corresponds to the simulated data in Fig. (3). From our derivation $\mu_f$ for this system should be 31 layers. Visually, we see that the bottom 31 layers, marked in gray, are indeed in a frozen regime, an area where particles do not switch position with other particles.

\newpage


\begin{figure}
\begin{center}
\epsfig{file=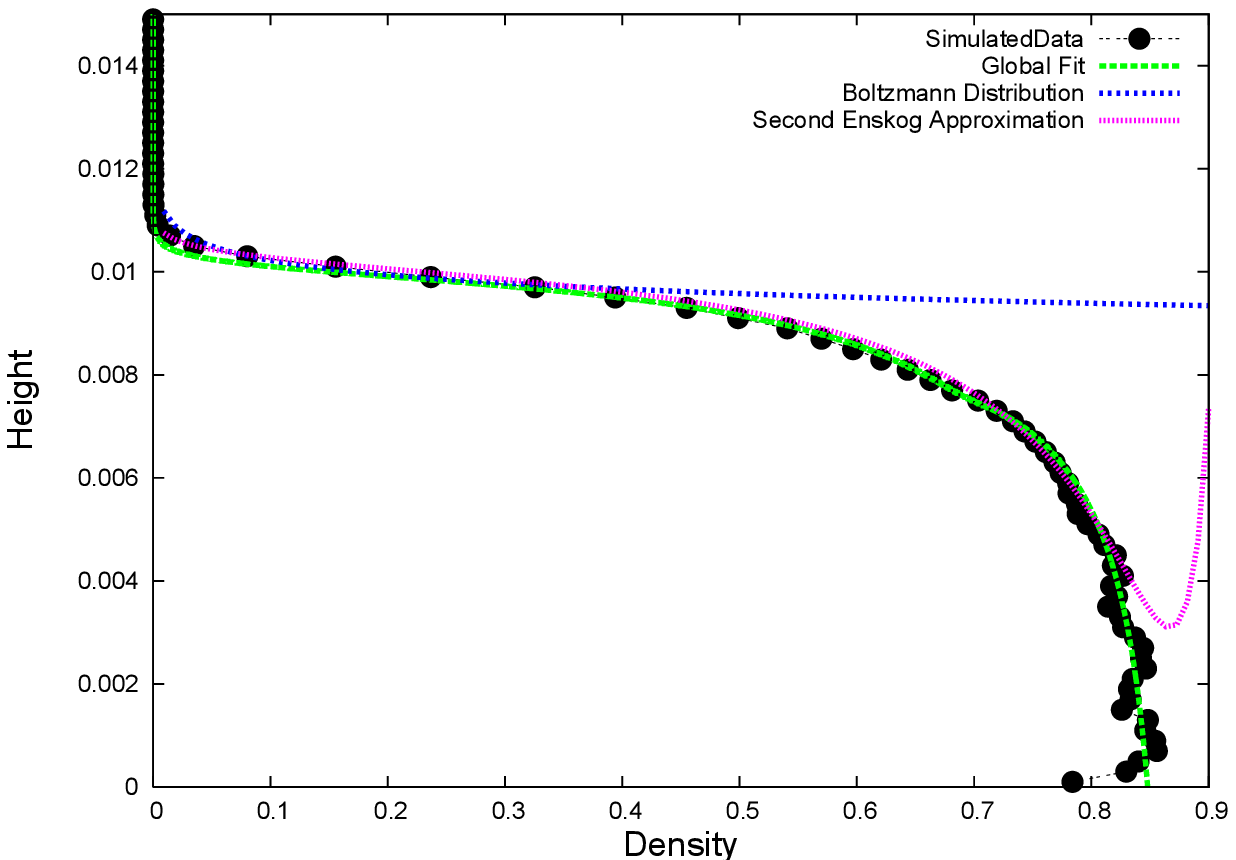, height=10cm, clip=}
\end{center}
\end{figure}

\begin{figure}
\begin{center}
\epsfig{file=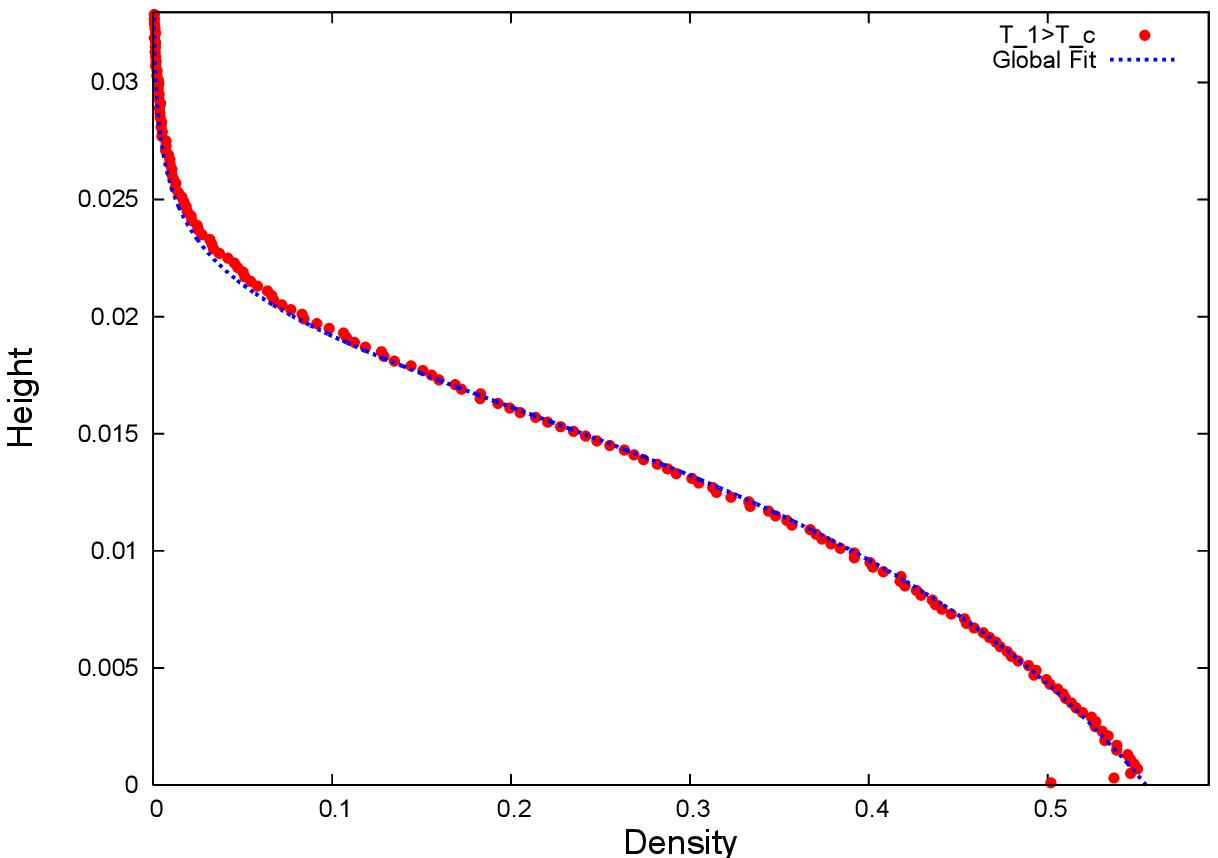, height=10cm, clip=}
\end{center}
\end{figure}

\begin{figure}
\begin{center}
\epsfig{file=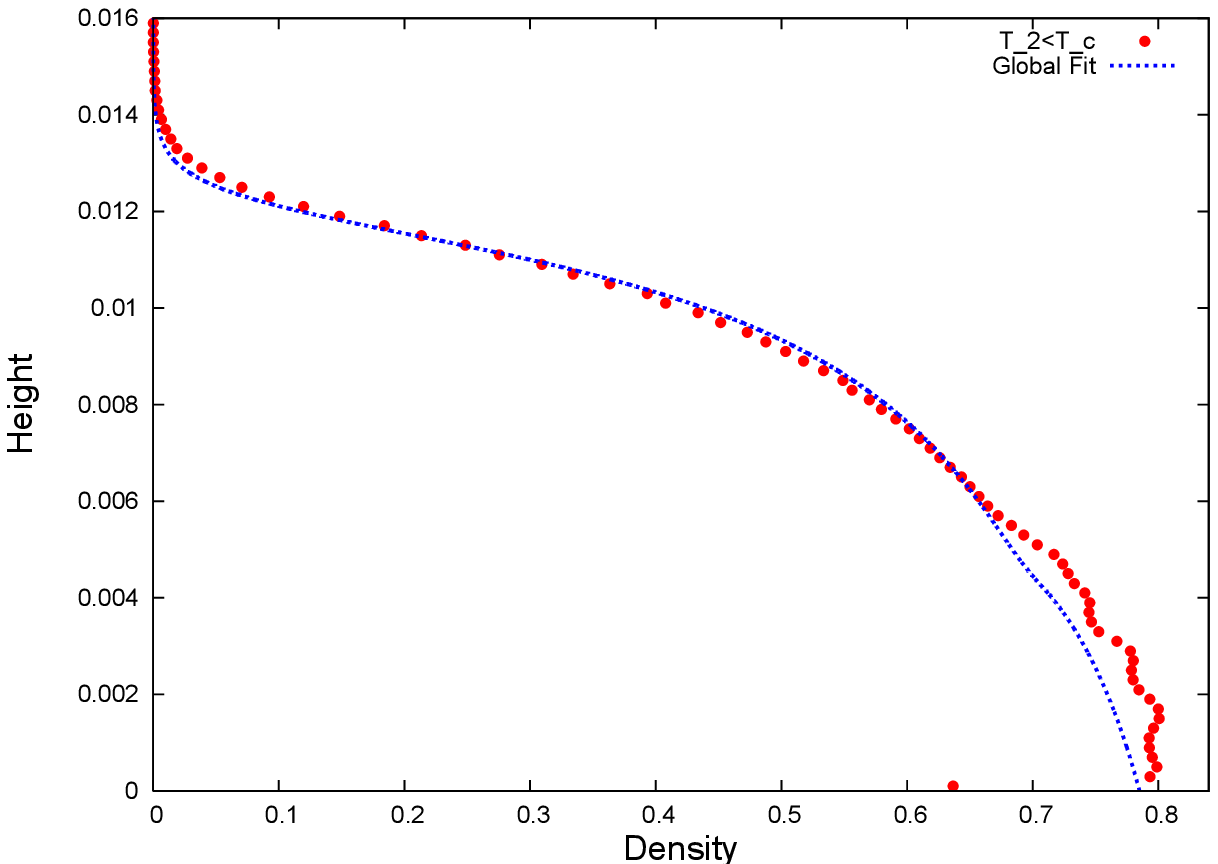, height=10cm, clip=}
\end{center}
\end{figure}

\begin{figure}
\begin{center}
\epsfig{file=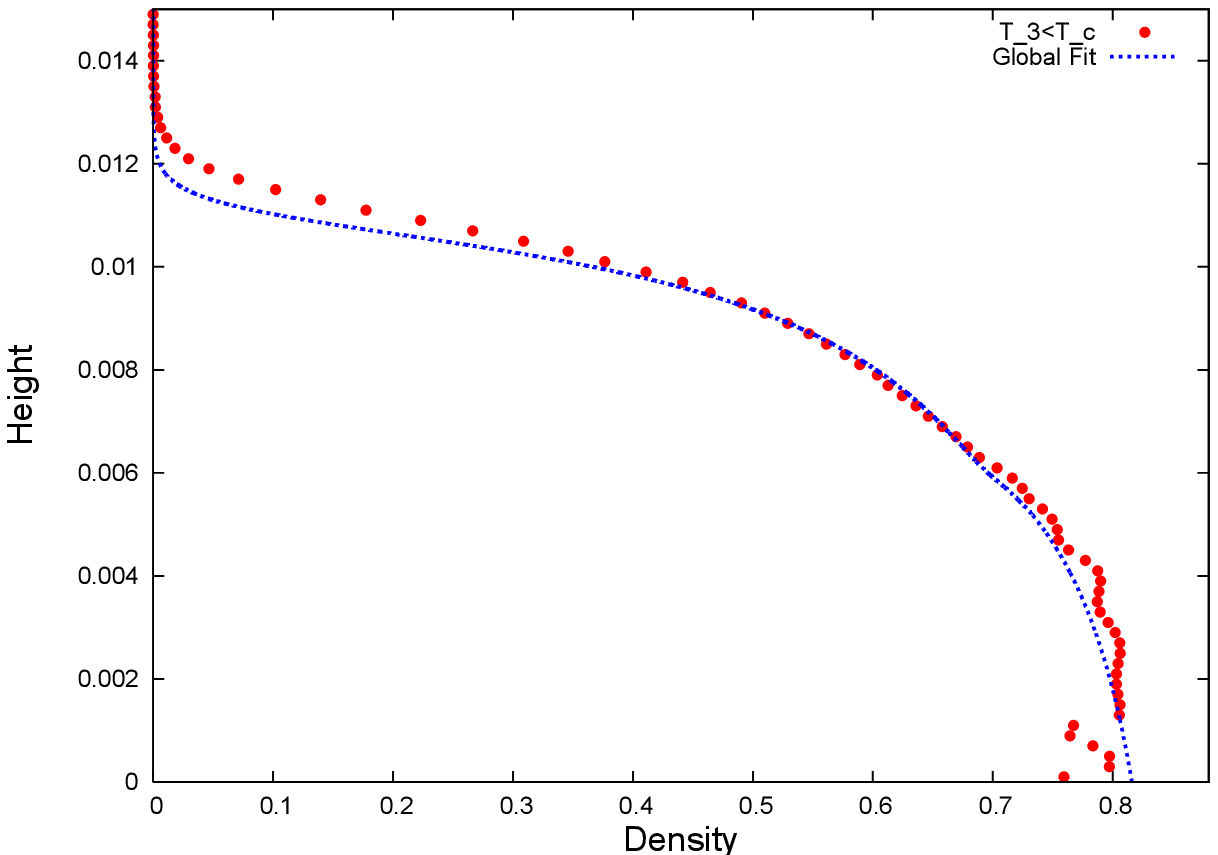, height=10cm, clip=}
\end{center}
\end{figure}

\begin{figure}
\begin{center}
\epsfig{file=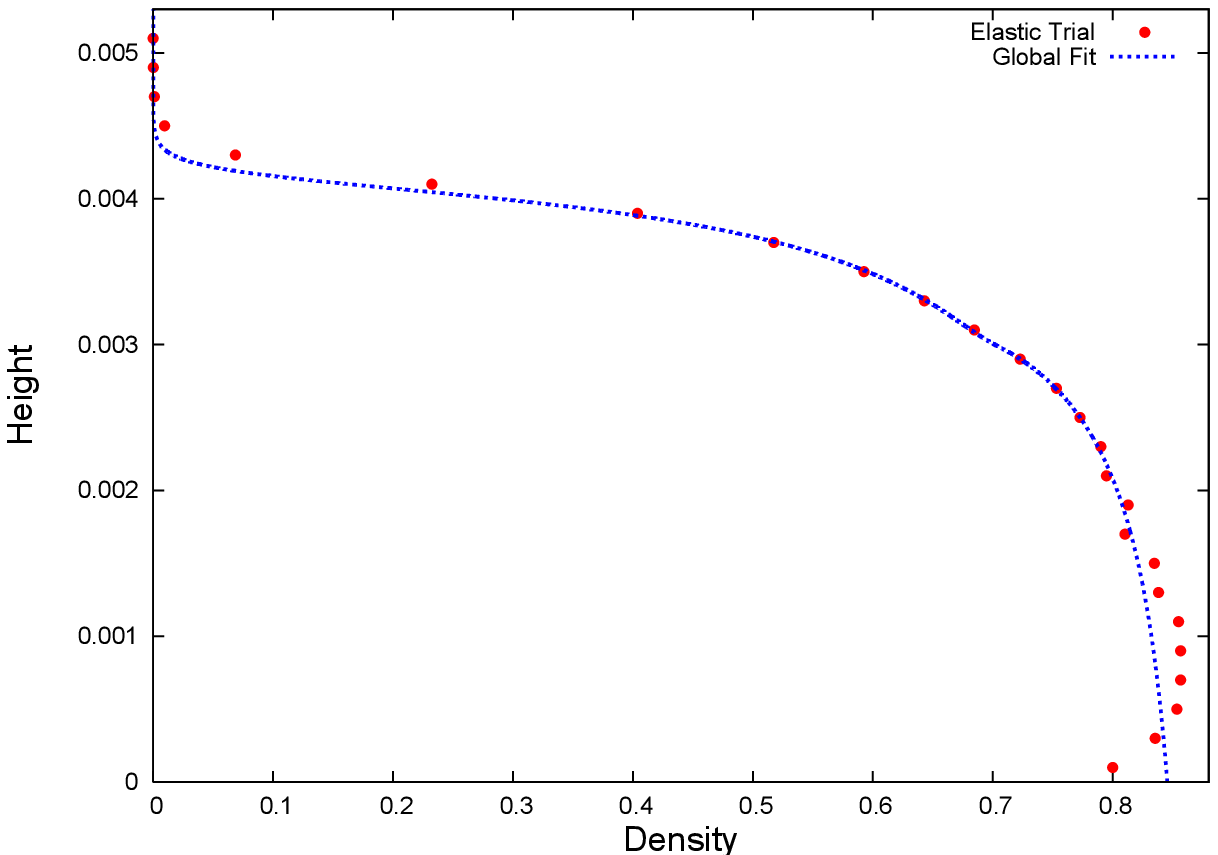, height=10cm, clip=}
\end{center}
\end{figure} 

\begin{figure}
\begin{center}
\epsfig{file=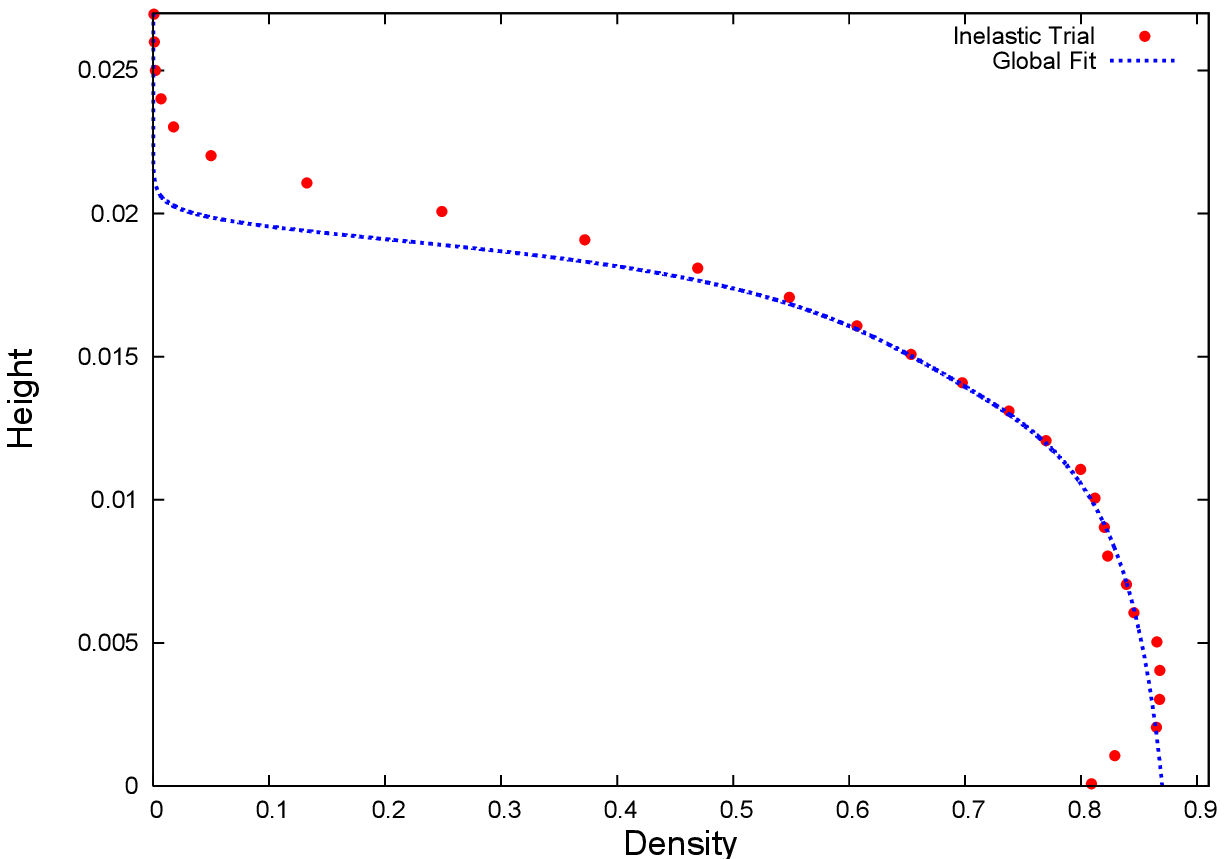, height=10cm, clip=}
\end{center}
\end{figure}

\begin{figure}
\begin{center}
\epsfig{file=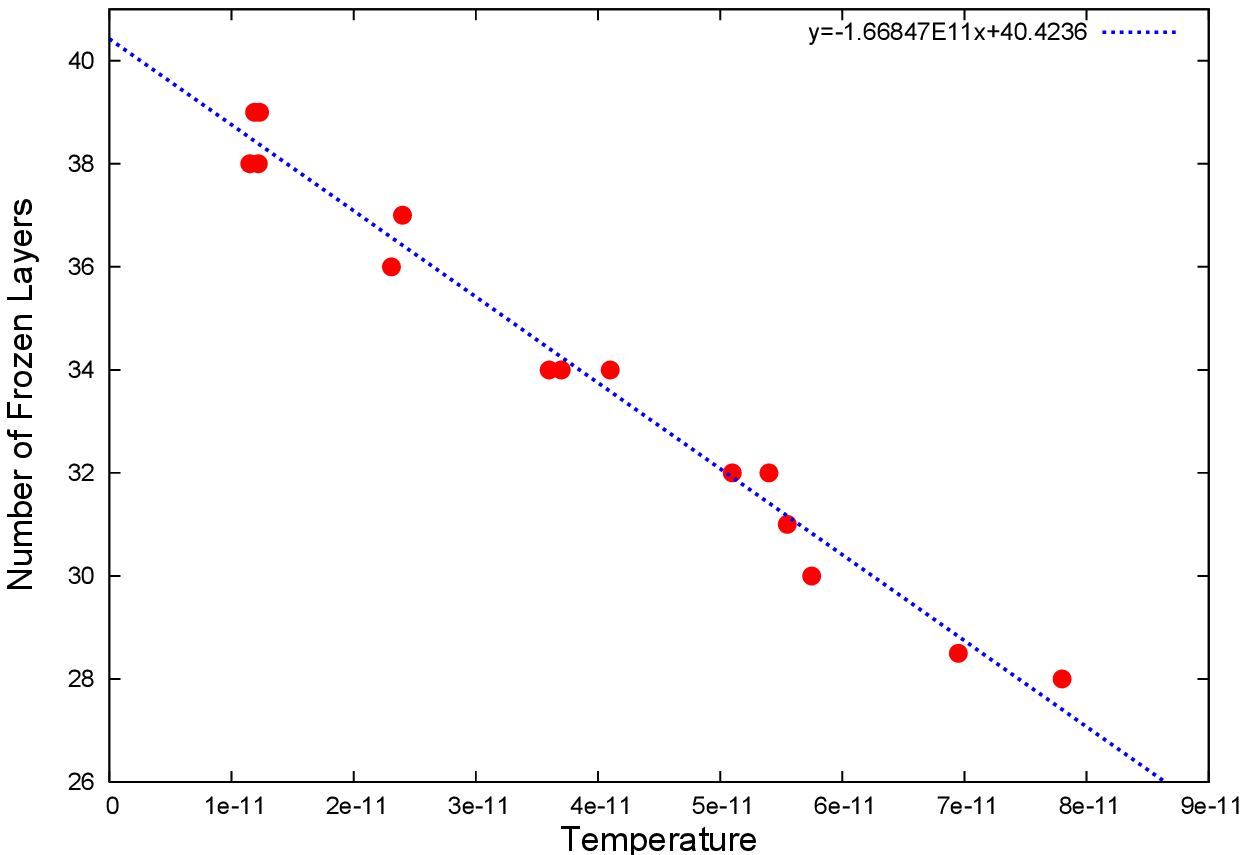, height=10cm, clip=}
\end{center}
\end{figure}



\end{document}